\documentclass[aps,prd,eqsecnum,nofootinbib,showpacs,twocolumn]{revtex4-1}

\usepackage{amsmath,amssymb,bm}
\usepackage[dvips]{graphicx}
\usepackage{color}

\newcommand{\be}{\begin{equation}}
\newcommand{\ee}{\end{equation}}
\newcommand{\bea}{\begin{eqnarray}}
\newcommand{\eea}{\end{eqnarray}}

\newcommand{\eq}[1]{Eq.~(\ref{eq:#1})}
\newcommand{\sect}[1]{Sec.~\ref{sec:#1}}
\newcommand{\appen}[1]{Appendix~\ref{sec:#1}}

\newcommand{\del}{\partial}

\newcommand{\Tc}{T_c}
\newcommand{\rh}{r_h}

\newcommand{\bra}{\langle}
\newcommand{\ket}{\rangle}
\newcommand{\calO}{{\cal O}}

\newcommand{\eg}{{\it e.g.}}
\newcommand{\ie}{{\it i.e.}}

\newcommand{\sourceA}[1]{A^{(0)}_{#1}}
\newcommand{\vevA}[1]{A^{(1)}_{#1}}

\newcommand{\bh}{black hole\ }

\newcommand{\SC}{superconductor}
\newcommand{\SCs}{superconductors}
\newcommand{\HSC}{holographic superconductor\ }
\newcommand{\HSCs}{holographic superconductors\ }
\newcommand{\LHSC}{holographic Lifshitz superconductor}
\newcommand{\LHSCs}{holographic Lifshitz superconductors}

\def\IR{\relax\text{I\kern-.18em R}}

\bmdefine{\bmq}{{\bm{q}}}
\bmdefine{\bmk}{{\bm{k}}}
\bmdefine{\bmx}{{\bm{x}}}
\bmdefine{\bmy}{{\bm{y}}}
\bmdefine{\bmr}{{\bm{r}}}
\bmdefine{\bmnabla}{{\bm{\nabla}}}
\bmdefine{\bmA}{ \bm{A} }
\bmdefine{\bmD}{ \bm{D} }
\newcommand{\bmF}{\bm{F}}
\bmdefine{\bmPhi}{ \bm{\Phi} }
\bmdefine{\bmPsi}{ \bm{\Psi} }
\bmdefine{\bmcalO}{ \bm{\mathcal{O}} }

\newcommand{\calF}{{\cal F}}

\newcommand{\calL}{{\cal L}}
\newcommand{\calM}{{\cal M}}

\newcommand{\epsmu}{\epsilon_{\mu}}

\newcommand{\zl}{z}
\newcommand{\zd}{z_\text{D}}

\newcommand{\nj}{\mathfrak{j}}

\newcommand{\tils}{\tilde{s}}
\newcommand{\tilu}{\tilde{u}}

\newcommand{\barmu}{\bar{\mu}}

\newcommand{\bmu}{\bar{\mu}}
\newcommand{\bepsmu}{\bar{\epsilon}_\mu}
\newcommand{\bw}{\bar{\omega}}
\newcommand{\bq}{\bar{q}}

\newcommand{\bky}{\bar{k}_y}
\newcommand{\bx}{\underline{x}}

\newcommand{\bay}{\bar{a}_y}
\newcommand{\bdel}{\bar{\del}}

\newcommand{\bbA}{\bar{\bmA}}
\newcommand{\bbB}{\bar{\bar{B}}}

\newcommand{\bbPsi}{\bar{\bmPsi}}

\newcommand{\deltas}{\delta_s}

\newcommand{\ctwo}{c_2}
\newcommand{\cfour}{c_4}
\newcommand{\cK}{c_K}
\newcommand{\cphi}{c_\phi}
\newcommand{\cA}{c_A}

\begin{document}


\title{Holographic Lifshitz superconductors: Analytic solution}
\author{Makoto Natsuume}
\email{makoto.natsuume@kek.jp}
\altaffiliation[Also at]{
Department of Particle and Nuclear Physics, 
SOKENDAI (The Graduate University for Advanced Studies), 1-1 Oho, 
Tsukuba, Ibaraki, 305-0801, Japan;
 Department of Physics Engineering, Mie University, 
 Tsu, 514-8507, Japan.}
\affiliation{KEK Theory Center, Institute of Particle and Nuclear Studies, 
High Energy Accelerator Research Organization,
Tsukuba, Ibaraki, 305-0801, Japan}
\author{Takashi Okamura}
\email{tokamura@kwansei.ac.jp}
\affiliation{Department of Physics, Kwansei Gakuin University,
Sanda, Hyogo, 669-1337, Japan}
\date{\today}
\begin{abstract}
We construct an analytic solution for a one-parameter family of holographic superconductors in asymptotically Lifshitz spacetimes. We utilize this solution to explore various properties of the systems such as  (1) the superfluid phase background and the grand canonical potential, (2) the order parameter response function or the susceptibility, (3) the London equation, (4) the background with a superfluid flow or a magnetic field. From these results, we identify the dual Ginzburg-Landau theory including numerical coefficients. Also, the dynamic critical exponent $\zd$ associated with the critical point is given by $\zd=2$ irrespective of the value of the Lifshitz exponent $\zl$. 
\end{abstract}
%

\maketitle

\section{Introduction and Summary}

The AdS/CFT duality \cite{Maldacena:1997re,Witten:1998qj,Witten:1998zw,Gubser:1998bc} has been a useful tool to study realistic strongly-coupled systems (see., \eg, Refs.~\cite{CasalderreySolana:2011us,Natsuume:2014sfa,Ammon:2015wua,Zaanen:2015oix,Hartnoll:2016apf} for textbooks). In condensed-matter applications, \HSCs provide particularly useful ``theoretical laboratories" \cite{Gubser:2008px,Hartnoll:2008vx,Hartnoll:2008kx}. They are useful to explore not only standard aspects of a superconducting transition but also various related phenomena such as critical dynamics \cite{Maeda:2009wv}, defect formations \cite{Sonner:2014tca,Chesler:2014gya,Natsuume:2017jmu}, and superfluid turbulences \cite{Adams:2012pj}.

The \HSCs arise in a broad range of gravitational theories with matter fields. From field theory point of view, this is natural since a \SC\ is a robust phenomenon at low temperature. For example, it arises not only in asymptotically AdS spacetimes but also in asymptotically Lifshitz spacetimes \cite{Brynjolfsson:2009ct,Sin:2009wi}, which is our main focus in this paper. 

A \HSC is typically an Einstein-Maxwell-complex scalar system. Such a system is hard to solve in general. One often needs either a numerical computation or an approximation method, and there are only a few analytic solutions \cite{Basu:2008bh,Herzog:2009ci,Herzog:2010vz}. In this paper, we present an analytic solution for a one-parameter family of \LHSCs.

A \LHSC\ has three parameters $(p,\zl,\Delta)$, where $p$ is the number of boundary spatial dimensions, $\zl$ is the Lifshitz exponent%
\footnote{The parameter $\zl$ is known as ``dynamic critical exponent," but we call it ``Lifshitz exponent" to avoid confusion with $\zd$ below.}, 
and $\Delta$ is the scaling dimension of the order parameter. 
In this paper, we consider the case where
\begin{enumerate}
\item $p=3\zl$, and
\item $\Delta=(p+\zl)/2$, or the scalar mass $m^2$ saturates the Lifshitz Breitenlohner-Freedman (BF) bound \cite{Breitenlohner:1982bm}.
\end{enumerate}
These conditions still admit a one-parameter family of theories, and these parameters are related by
\be
p=3\zl~, \Delta=2\zl~.
%
\ee
In this case, there is a simple analytic solution for the order parameter at the critical point: 
\be
\Psi=\frac{u^{2\zl}}{1+u^{2\zl}}~,
\label{eq:analytic_sol}
\ee
where $u$ is the inverse of the radial coordinate. The $\zl=1$ case, namely the AdS case, has been discussed in Ref.~\cite{Herzog:2010vz}. Below we utilize this solution to explore various properties of these systems, \eg, 
\begin{enumerate}

\item The background solution in the superfluid phase (\sect{bg_low}). 

\item The grand canonical potential, and the phase transition is second order (\sect{grand_canonical}).

\item The order parameter response function or the susceptibility. The dynamic response is obtained in the normal phase (\sect{high}), and the thermodynamic response is obtained in the superfluid phase (\sect{low_chi}). 

\item The London equation. From the equation, one gets the London penetration depth and the imaginary part of the conductivity has the $1/\omega$-pole which implies the diverging DC conductivity (\sect{vector}).

\item The background solution with a vector potential. This gives the critical superfluid velocity and the critical magnetic field (\sect{bg_vector}).

\item From these results, all static critical exponents $(\alpha,\beta,\gamma,\delta,\nu,\eta)$, the dynamic  critical exponent $\zd$, and the ratio of critical amplitudes. These results are consistent with the standard Ginzburg-Landau (GL) theory or the $\phi^4$ mean-field theory. We identify the dual GL theory including numerical coefficients (\sect{GL}).

\end{enumerate}

The \LHSCs\ have been studied previously, but it is still nice to analyze these properties all at once analytically for an infinite number of theories. First, in previous works, the system was studied mostly using numerical methods. Second, the system was studied only for some specific values of $(p,\zl,\Delta)$%
\footnote{
For example, Ref.~\cite{Brynjolfsson:2009ct} considers the $(p,\zl)=(2,3/2)$ case and obtains a charged Lifshitz black hole with scalar hair. The other works typically take the probe limit to study the system. Ref.~\cite{Sin:2009wi} considers the $(p, \zl, \Delta)=(2,2,3)$ case in a Lifshitz \bh background \cite{Balasubramanian:2009rx}. Ref.~\cite{Bu:2012zzb} considers the $(p, \zl, \Delta)=(2,2,3), (2,2,4)$ cases in the same Lifshitz \bh background as ours. Ref.~\cite{Lu:2013tza} considers $(p, \zl, \Delta)=(2,2,2), (2,2,3), (3,2,3), (3,3,3)$ cases in the same Lifshitz \bh background as ours. As far as we are aware, our $(p,\zl,\Delta)=(3\zl,\zl,2\zl)$ case was not studied before. 
}.
Third, some of the above properties were studied but not all were studied.

In particular, previous works typically have shown that (i) there exists a $\Psi\neq0$ solution at low temperatures, (ii) the solution is favorable from the free energy or from the grand canonical potential, (iii) the spontaneous condensate has the standard $\phi^4$ mean-field exponent $\beta=1/2$, and (iv) the diverging DC conductivity. 

On the other hand, the other properties are newly investigated, \eg, the other critical exponents as well as exact expressions for various numerical coefficients including critical amplitudes. Also, the critical dynamics of a  \LHSC\ has never been investigated%
\footnote{Dynamics in Lifshitz geometry has been studied, \eg, in Refs.~\cite{Sybesma:2015oha,Keranen:2016ija}. Critical dynamics has been studied in holography, \eg, in Refs.~\cite{Maeda:2008hn,Maeda:2009wv,Buchel:2010gd,Natsuume:2010bs}.}. 
At a finite-temperature critical point, the correlation length $\xi$ and the relaxation time $\tau$ of the order parameter obey a scaling law:
\be
\tau \propto \xi^{\zd}~.
\label{eq:scaling_DCP}
\ee
We obtain $\zd=2$ irrespective of the value of the Lifshitz exponent $\zl$. We discuss the relation between $\zl$ and $\zd$ in \sect{two_z}.

\section{Preliminaries}

\subsection{Lifshitz \bh}\label{sec:EMD}

The Lifshitz geometry \cite{Kachru:2008yh} is given by
\be
ds_{p+2}^2 = -\left(\frac{r}{L}\right)^{2\zl} dt^2 + \left(\frac{r}{L}\right)^{2} dx_i^2 + L^2 \frac{dr^2}{r^2}
%
\ee
 (see, \eg, Ref.~\cite{Taylor:2015glc} for a review). The geometry is invariant under an anisotropic scaling
\be
t \to a^{\zl} t~, \quad
x^i \to a x^i~,\quad 
r \to r/a~.
\label{eq:scaling_Lifshitz}
\ee
There are various Lifshitz \bh solutions known in the literature, both analytically and numerically, depending on bulk theories. We use the solution in Refs.~\cite{Taylor:2008tg,Pang:2009ad}. The metric is given by
\begin{subequations}
\label{eq:Lifshitz_BH}
\begin{align}
ds^2
 &= -\left(\frac{r}{L}\right)^{2\zl} h dt^2 + \left(\frac{r}{L}\right)^{2} dx_i^2 + L^2 \frac{dr^2}{r^2h}~, \\
&= -\left(\frac{\rh}{L}\right)^{2\zl} \frac{h}{u^{2\zl}} dt^2 + \left(\frac{\rh}{L}\right)^{2} \frac{dx_i^2}{u^2} + L^2\frac{du^2}{u^2h}~, \\
h &= 1- \left(\frac{\rh}{r}\right)^{p+\zl} = 1- u^{p+\zl}~,
%
\end{align}
\end{subequations}
where $u:=\rh/r$, and $\rh$ is the horizon radius. The metric is invariant under the Lifshitz scaling \eqref{eq:scaling_Lifshitz} with the scaled horizon radius $\rh\to\rh/a$. The Hawking temperature is given by
\be
T = \frac{p+\zl}{4\pi L} \left(\frac{\rh}{L} \right)^{\zl}~. 
%
\ee

The metric can be obtained as a solution of an Einstein-Maxwell-dilaton system%
\footnote{We use capital Latin indices $M, N, \ldots$ for the $(p+2)$-dimensional bulk spacetime coordinates and use Greek indices $\mu, \nu, \ldots$ for the $(p+1)$-dimensional boundary coordinates. The boundary coordinates are written as  $x^\mu = (t, x^i)=(t,x,y,\cdots)$.}:
\begin{align}
  S = \frac{1}{16\pi G_{p+2}} &
  \int d^{p+2}x\, \sqrt{-g} \bigg\{ 
  R - 2\Lambda 
\nonumber \\ &
  - \frac{1}{2} (\del_M\phi)^2 - \frac{1}{4} e^{\lambda \phi}\calF_{MN}^2 
  \bigg\}~,
\label{eq:EMD}
%
\end{align}
where $G_{p+2}$ is the $(p+2)$-dimensional Newton's constant and
\begin{subequations}
\begin{align}
&\Lambda = - \frac{(p+\zl-1)(p+\zl)}{2L^2}~, \quad
\lambda^2 = \frac{2p}{\zl-1}~, \\
&\calF_{MN} = 2\, \partial_{[M} {\cal A}_{N]}~.
\end{align}
\end{subequations}
The matter field solutions are given by
\begin{subequations}
\begin{align}
e^{\lambda \phi} &= u^{2p}~, \\
{\cal A}_t &= -\sqrt{ \frac{2(\zl-1)}{p+\zl} } \frac{1}{u^{p+\zl}}~.
%
\end{align}
\end{subequations}
But for our purpose, the point using this solution is that (i) it provides an analytic Lifshitz \bh solution, and (ii) a class of \HSCs in this background admits an analytic solution.

\subsection{Holographic Lifshitz \SCs}

We couple an additional matter system, a Maxwell-complex scalar system in addition to the above system  \cite{Brynjolfsson:2009ct,Sin:2009wi}:
\be
  S = -\frac{1}{e^2} 
  \int d^{p+2}x\, \sqrt{-g} \left\{ 
  \frac{1}{4} F_{MN}^2 + \left\vert D_M \Psi \right\vert^2 + V(\Psi) 
  \right\}~,
\label{eq:Lifshitz_HSC}
\ee
where
\begin{align}
  & F_{MN} = 2\, \partial_{[M} A_{N]}~,
& & D_M := \nabla_M - i A_M~,
& & V= m^2 |\Psi|^2~.
\end{align}
The $U(1)$-field $A_M$ is different from ${\cal A}_M$ in \eq{EMD}.

We take the probe limit $e \gg 1$, where the backreaction of these matter fields onto the geometry is ignored. Namely, we solve the system \eqref{eq:Lifshitz_HSC} in the background \eqref{eq:Lifshitz_BH}. The equations of motion are given by 
\begin{subequations}
\label{eq:eom}
\begin{align}
(D^2-m^2)\Psi &= 0~, \\
\nabla_N F^{MN} &= j^M 
\\
&:=ig^{MN}[(D_N\Psi)^\dag\Psi- \Psi^\dag(D_N\Psi)]~. 
%
\end{align}
\end{subequations}
In the $A_u=0$ gauge, the asymptotic behaviors of the matter fields are given by
\begin{subequations}
\begin{align}
A_t & \sim A_t^{(0)}  + A_t^{(1)}\tilu^{p-\zl}~, \qquad (p>\zl)~, \\
A_i & \sim A_i^{(0)}  + A_i^{(1)} \tilu^{p+\zl-2}~, \\
\Psi &\sim \Psi^{(0)} \tilu^{\Delta_-} + \Psi^{(1)} \tilu^{\Delta_+}~, \\
\tilu &:=\frac{L}{r}~, \\
\Delta_\pm &:= \frac{p+\zl}{2} \pm \sqrt{ \frac{(p+\zl)^2}{4} + L^2m^2 }~,
\end{align}
\end{subequations}
where $A_t^{(1)}$ represents the charge density $\rho$, and $A_t^{(0)}=\mu$ is the chemical potential. Similarly, $A_i^{(1)}$ represents the current density $ J^i$, and $A_i^{(0)}$ is the vector potential. For $\Psi$, $\Psi^{(1)}$ represents the order parameter $\calO$, and $\Psi^{(0)}$ is the external source for $\calO$. (See \appen{dict} for the precise dictionary.)
Then, the BF bound in the asymptotically Lifshitz geometry is given by
\be
m_\text{BF}^2 = - \frac{(p+\zl)^2}{4L^2}~.
%
\ee
When the BF bound is saturated, the asymptotic behavior is replaced by
\begin{align}
\Psi &\sim \Psi^{(0)} \tilu^\Delta \ln \tilu +\Psi^{(1)} \tilu^\Delta~, \quad 
\Delta := \frac{p+\zl}{2}~.
%
\end{align}

The equations of motion \eqref{eq:eom} admit a solution
\begin{subequations}
\label{eq:sol_high}
\begin{align}
\bmA_t &=  \mu(1-u^{p-\zl})~, \quad (p>\zl)~, \\
\bmA_i &= \bmA_u = 0~, \\
\bmPsi &=0~,
%
\end{align}
\end{subequations}
where boldface letters indicate background values. But, at the critical point, the $\bmPsi=0$ solution becomes unstable and is replaced by a $\bmPsi\neq0$ solution. We see this in detail below.

\section{Critical point}\label{sec:critical}

Below we consider the case $p=3\zl$. 
It is convenient to introduce a new coordinate $s:=u^{2\zl}$. The metric then becomes%
\footnote{It is well-known that the $\zl=\infty$ limit of the Lifshitz geometry is AdS$_2 \times \IR^p$ asymptotically. But this is the case for a finite $p$. In our case, $p=3\zl$, and the $\zl=\infty$ limit does not reduce to AdS$_2$.}
\begin{subequations}
\begin{align}
ds^2
 &= - \left(\frac{\rh}{L}\right)^{2\zl} \frac{h}{s} dt^2 + \left(\frac{\rh}{L}\right)^2 \frac{dx_i^2}{s^{1/\zl}} + \left(\frac{L}{2\zl}\right)^2\frac{ds^2}{s^2h}~, 
\\
h &= 1- s^2~.
%
\end{align}
\end{subequations}

We consider the scalar which saturates the Lifshitz BF bound $m^2=-(2\zl/L)^2$. The scaling dimension $\Delta$ is given by $\Delta=2\zl$. First, consider the static homogeneous solution $\bmPsi=\bmPsi(s)$, and approach the critical point from high temperature. Near the critical point, the scalar field $\bmPsi$ remains small, and one can ignore the backreaction of $\bmPsi$ onto the Maxwell field. In this region, one can use \eq{sol_high} for the Maxwell field, and it is enough to solve the $\bmPsi$-equation%
\footnote{We later use a perturbative expansion for a systematic analysis (\sect{bg_low}).}.
The $\bmPsi$-equation becomes
\begin{align}
&\del_s \left(\frac{h}{s}\del_s \bmPsi \right) + \left\{ \left(\frac{\mu}{2\pi T}\right)^2\frac{(1-s)^2}{hs^2} + \frac{1}{s^3} \right\}\bmPsi =0~, 
\label{eq:eom_critical} \\
&T = \frac{\zl}{\pi L} \left(\frac{\rh}{L} \right)^{\zl}~.
%
\end{align}
Thus, the solution is parametrized by a dimensionless parameter $\mu/T$. The equation admits a solution
\begin{align}
\bmPsi \propto \frac{s}{1+s}~, 
\quad 
&\text{at }  \left(\frac{\mu}{T}\right)_c=2\pi~, 
\label{eq:sol_critical} \\
&\text{or } \frac{\mu}{\frac{1}{L}(\frac{\rh}{L})^{\zl}}=2\zl~.
\nonumber 
\end{align}
This is the solution at the critical point. 

The $\zl$-dependence disappears in \eq{eom_critical}, and it only appears implicitly in the definition of $T$. One can understand this as follows. For the static homogeneous solution, the Laplacian becomes
\be
\nabla^2\bmPsi = \frac{1}{\sqrt{-g}}\del_s(\sqrt{-g}g^{ss}\del_s\bmPsi)~.
%
\ee
The $\zl$-dependence appears in the boundary spatial metric $g_{ij}$, and it appears only through $\det g$. But in our case,
\begin{align}
-\det g \propto s^{-3-p/\zl}=s^{-6}~,
%
\end{align}
so $p$ and $\zl$ disappear. The metric $g^{ss}$ is also proportional to $\zl^2$, but it is factored out in the $\bmPsi$-equation since $m^2 \propto -\zl^2$ and $1/T^2 \propto 1/\zl^2$. It then follows that the $\bmPsi$-equation formally reduces to the same equation for all $\zl$. 

Note that \eq{sol_critical} is the solution directly at the critical point. As one lowers temperature further, the solution is modified, and we construct the background solution $\bmPsi, \bmA_t$ in \sect{low}. The $\zl$-dependence can also be eliminated from the $\bmA_t$-equation [by redefining $\bmPsi$ and $\bmA_t$ as in \eq{redefine}.] Thus, the static homogeneous solution is essentially the same as the $\zl=1$ case apart from various factors of $(2\zl)$. Then, from the analysis of Ref.~\cite{Herzog:2010vz}, \eq{sol_critical} is the solution at the critical point, and the solution has a lower grand canonical potential than the $\bmPsi=0$ solution at low temperature. 

However, the full equations of motion do not reduce to the same equations as the $\zl=1$ case. In general, more nontrivial $\zl$-dependences appear. For example, they appear when one considers 
\begin{itemize}
\item inhomogeneous perturbations in the boundary spatial directions \eg, $\delta\phi \propto e^{iqx} $ (\sect{high}), or
\item perturbations or solutions with a vector potential $A_i$ (\sect{vector} and \sect{bg_vector}).
\end{itemize}

Below we construct the background solution $\bmPsi, \bmA_M$. We also consider the linear perturbations from the background:
\begin{subequations}
\begin{align}
\Psi &=\bmPsi+\delta\Psi~, \\
A_M &= \bmA_M+\delta A_M~.
%
\end{align}
\end{subequations}
We take the gauge $\bmA_s=\delta A_s=0$.
We consider the perturbations of the form 
\be
\delta\phi(k) \sim e^{-i\omega t+iqx}~,
%
\ee
where $k^\mu=(\omega, q, 0, \cdots)$. Then, the Maxwell perturbations are decomposed as 
\begin{itemize}
\item vector modes, \eg, $\delta A_y$, and 
\item scalar modes $\delta A_t, \delta A_x$ which can couple to $\delta\Psi$ in general.
\end{itemize}
For simplicity, we set $e=L=\rh=1$ below. In this unit, $\mu_c=2\zl$, and we vary the chemical potential $\mu$. Also, we often use quantities with ``$~\bar{~}~$". All quantities with ``$~\bar{~}~$" are defined by
\be
\bar{\phi}:=\frac{\phi}{2\zl}~,
\label{eq:redefine}
\ee
when $\rh=1$. For example, $\bmu_c=1$. We restore units for some of our main results in \appen{restore}.

\section{High-temperature phase}\label{sec:high}

At high temperature, the background solution is given by
\begin{subequations}
\begin{align}
\bmA_t &= \mu(1-s)~, \\
\bmA_i &=0~, \\
\bmPsi &=0~.
%
\end{align}
\end{subequations}

The interesting quantity in the high-temperature phase is the ``order parameter response function," the susceptibility, or the correlation function of the order parameter. We show that the response function takes the form
\begin{subequations}
\label{eq:response_fn}
\begin{align}
   \chi_{k}
  & = \frac{ \delta\calO(k) }{ \delta\Psi^{(0)}(k) }
\nonumber \\
  &\propto
    \frac{1}{- \frac{2i}{\cK\Gamma} \omega + q^2 + \frac{1}{\xi^2} }~,
\\
    \xi^2 &\propto |\epsmu|^{-1}~,
\label{eq:correl_length}
\end{align}
\end{subequations}
for a small $(\omega, q, \epsmu)$, where $\epsmu:=\mu-\mu_c$, and $\cK$ and $\Gamma$ are parameters we use to compare with the GL theory (\sect{GL}). The function gives the following information:
\begin{itemize}
\item
The $\omega=q=0$ limit is the thermodynamic response function 
\begin{subequations}
\begin{align}
\chi_T=A/|\epsmu|~,
%
\end{align}
where the coefficient $A$ is known as the critical amplitude. Then, the exponent $\gamma$ defined by 
$
\chi_T\propto |\epsmu|^{-\gamma}
$ 
is $\gamma=1$. 

\item
The $\omega=0$ limit is the static response 
\begin{align}
\chi_{\omega=0, q} \propto (q^2+\xi^{-2})^{-1}~.
%
\end{align}
Then, $\xi$ is the correlation length, and the exponent $\nu$ defined by 
$
\xi \propto |\epsmu|^{-\nu}
%
$
is $\nu=1/2$ from \eq{correl_length}. Also, the anomalous exponent $\eta$ defined by 
$
\chi_{\omega=0, q}|_{\mu_c} \propto q^{-2+\eta}
%
$
is $\eta=0$.  
\item 
The $\omega\neq0$ case is the dynamic response. Then, the relaxation time behaves as 
\begin{align}
\tau_{q=0} \propto \xi^2~,
%
\end{align}
and the dynamic critical exponent $\zd$ defined by 
$
\tau_{q=0} \propto \xi^{\zd}
%
$
is $\zd=2$.
\end{subequations}
\end{itemize}
Thus, the computation determines the exponents $(\gamma,\nu,\eta,\zd)$ as well as the critical amplitude $A$. An explicit solution is not really necessary to compute critical exponents, and analytic arguments are possible \cite{Maeda:2009wv,Natsuume:2017jmu}. On the other hand, an explicit solution is useful to obtain various numerical coefficients such as $A$.

The response function can be obtained from the bulk scalar field $\Psi$. 
Consider the linear perturbation from the background $\Psi=\bmPsi+\delta\Psi$. From the bulk point of view, the response function pole corresponds to a quasinormal pole of $\delta\Psi$. When $\bmPsi=0$, Maxwell scalar modes $\delta A_t$ and $\delta A_x$ decouple from the $\delta\Psi$-equation%
\footnote{
The Maxwell scalar modes give a diffusion pole, and one can determine the diffusion constant. But at high temperature, the computation is not unique to holographic superconductors. It is just a Maxwell field problem in the Lifshitz background. The Maxwell vector mode can determine the conductivity. Again, at high temperature, the computation is not unique to holographic superconductors. But we compute the vector mode in the low-temperature phase. The $O(\omega)$-coefficient is common both to the high-temperature and the low-temperature phases.
}. 
Thus, to determine the order parameter response, it is enough to consider the $\delta\Psi$-equation:
\begin{align}
\del_s \left(\frac{h}{s}\del_s \delta\Psi \right) 
+ \left\{ \frac{(\bbA_t+\bw)^2}{hs^2} -\frac{\bq^2}{s^{3-1/\zl}} + \frac{1}{s^3} \right\}\delta\Psi = 0~. 
%
\end{align}
Asymptotically, we impose the boundary condition $\delta\Psi(s\to0)=\delta\Psi^{(0)} s \ln s/(2\zl)$. At the horizon, we impose the incoming-wave boundary condition. 

The $\delta\Psi$ perturbation cannot be solved for a generic $\mu$, so we set $\bepsmu=\bmu-1<0$ and employ the $\bepsmu$-expansion as well as the $(\omega,q)$-expansion:
\begin{align}
\delta\Psi(s,k)&=(1-s)^{-i\bw/2}
\\ &\quad \times %
((\psi_c + \bepsmu \psi_\epsilon +\cdots)+ \bw \psi_\omega + \bq^2 \psi_q +\cdots)~.
\nonumber
%
\end{align}
This form is taken to implement the incoming-wave boundary condition. Then, the boundary condition reduces to the regularity condition for $\psi_c$ and so on. The equation of motion reduces to
\begin{subequations}
\begin{align}
\calL_\psi \psi_c &= 0~, \\
\calL_\psi \psi_i &= \nj_i(\psi_c)~,
%
\end{align}
where
\be
\calL_\psi = \del_s \left(\frac{h}{s}\del_s \right) + \left\{ \frac{(1-s)^2}{hs^2} + \frac{1}{s^3} \right\}~,
%
\ee
\end{subequations}
and the index $i$ collectively represents $\epsilon, \omega$, and $q$. The homogeneous equation $\calL_\psi \psi_c =0$ can be solved as
\begin{align}
\psi_c &= c_1 \frac{s}{1+s} + c_2 \frac{s}{1+s} \ln\left[ \frac{s}{(1-s)^2}\right]~. 
%
\end{align}
From the regularity at the horizon, $c_2=0$.

The source terms of inhomogeneous equations then become
\begin{subequations}
\begin{align}
\nj_\epsilon &= -\frac{2(1-s)}{s(1+s)^2} c_1~, \\
\nj_\omega &=\frac{-4-i(1+s)}{2s(1+s)^2} c_1~, \\
\nj_q &= \frac{1}{s^{2-1/\zl}(1+s)} c_1~.
%
\end{align}
\end{subequations}
The $\psi_\epsilon$ and $\psi_\omega$ solutions are
\begin{subequations}
\begin{align}
\psi_\epsilon &= - c_1\frac{s}{1+s} \left\{ -\frac{1}{2}\ln s+\ln (1+s)\right\} 
\\
& \sim c_1 \frac{1}{2} s\ln s \quad (s\to 0)~,
\\
\psi_\omega &= c_1 \frac{s}{4(1+s)} \left\{ (3+i)\ln s-2\ln (1+s)\right\}
\\
& \sim c_1 \frac{3+i}{4} s\ln s \quad (s\to 0)~.
%
\end{align}
\end{subequations}
For $\psi_q$, we discuss the $\zl=1$ and $\zl\neq1$ cases separately.

\subsection{$\zl=1$}

The $\psi_q$ solution is given by
\begin{align}
\psi_q &= -c_1\frac{1}{2(1+s)} s\ln s
\\
& \sim -c_1 \frac{1}{2} s\ln s \quad (s\to 0)~.
%
\end{align}
The asymptotic behavior then becomes
\begin{align}
\delta\Psi &\sim c_1(1-s)^{-i\bw/2} 
 \\ & \times %
\left\{-\frac{1}{2}\left(\bq^2 - \frac{3+i}{2} \bw + |\bepsmu| +\cdots \right) s\ln s + s +\cdots \right\}~.
\nonumber %
%
\end{align}
The asymptotic boundary condition determines $c_1$. Then, the order parameter response function becomes
\begin{align}
\chi_{\omega, q}^>
&= \frac{ \delta\calO }{ \delta\Psi^{(0)} } \nonumber\\
&= \frac{2}{ \frac{1}{2} q^2 - \frac{3+i}{2} \omega + |\epsmu| }~,
%
\end{align}
where we use the dictionary in \appen{dict}. The response function indeed takes the form of \eq{response_fn}, and 
\begin{subequations}
\begin{align}
\xi_>^2 &= \frac{1}{2|\epsmu|}~, \\
A_> &= 2~.
%
\end{align}
\end{subequations}
The dispersion relation is given by
\be
\omega = \frac{3-i}{5}\left( |\epsmu|+\frac{1}{2}q^2+\cdots \right)~.
%
\ee
The relaxation time $\tau$ then becomes
\be
\tau^{-1} = \frac{1}{5}|\epsmu|=\frac{1}{10}\xi_>^{-2}~.
%
\ee

\subsection{$\zl>1$}
In this case, we are not able to obtain the generic expression for $\psi_q$. Besides, even when the analytic expression is available, it is too cumbersome to write here. However, the slow falloff has a simple expression:
\begin{subequations}
\begin{align}
\psi^{(0)}_q &= - c_1 I(\zl)~, \\
I(\zl) &:=\int_0^1 \frac{ds}{s^{1-1/\zl}(1+s)^2} \\
&= \frac{1}{2} + \frac{1}{2}\left(1-\frac{1}{\zl}\right) 
\nonumber \\
&\quad\times\left\{ \psi_0\left(\frac{1}{2\zl}+\frac{1}{2} \right) - \psi_0\left(\frac{1}{2\zl}\right) \right\}~,
%
\end{align}
\end{subequations}
where $\psi_0(x)$ is the digamma function:
\begin{align}
\psi_0(x) &= \frac{d}{dx}\ln \Gamma(x)~.
%
\end{align}
A few examples of $I(\zl)$ are
\begin{align}
I(1) &= \frac{1}{2}~, 
\nonumber \\
I(2) &= \frac{1}{2} +\frac{\pi}{4} \approx 1.285~, 
\nonumber \\
I(3) &= \frac{1}{2} +\frac{2}{3}\left(\ln2+\frac{\pi}{\sqrt{3}} \right) \approx 2.171~.
\nonumber 
%
\end{align}
The combination $I(\zl)/\zl$ monotonically increases with $\zl$ and reaches 1 for $\zl\to\infty$. 
In order to obtain the falloff, we essentially used the standard method to solve an inhomogeneous differential equation (\appen{Green}).

Then, the order parameter response function becomes
\begin{align}
\chi_{\omega, q}^>
= \frac{2}{ \frac{I(\zl)}{\zl} q^2 - \frac{3+i}{2} \omega + |\epsmu|  }~,
\label{eq:response_holo}
\end{align}
which gives
\begin{subequations}
\begin{align}
\xi_>^2 &= \frac{I(\zl)/\zl}{|\epsmu|}~, \\
A_> &= 2~.
%
\end{align}
\end{subequations}
The dispersion relation is given by
\be
\omega = \frac{3-i}{5}\left( |\epsmu|+\frac{I(\zl)}{\zl}q^2+\cdots \right)~.
%
\ee
The relaxation time is given by
\be
\tau^{-1} = \frac{1}{5}|\epsmu| = \frac{I(\zl)}{5\zl}\xi_>^{-2}~.
%
\ee

\subsection{$\zl\gg1$}
For large $\zl$, 
\be
I(\zl) 
\approx \zl~,
%
\ee
so 
\begin{subequations}
\begin{align}
\chi_{\omega, q}^> &\approx \frac{2}{ q^2 - \frac{3+i}{2} \omega + |\epsmu| }~, \\
\omega &\approx \frac{3-i}{5}(|\epsmu|+q^2+\cdots)~, \\
\xi_>^2 &\approx \frac{1}{|\epsmu|}~, \quad
\tau^{-1} = \frac{1}{5}|\epsmu|\approx\frac{1}{5}\xi_>^{-2}~.
%
\end{align}
\end{subequations}

\section{Low-temperature phase}\label{sec:low}

In the low-temperature phase, our task is
\begin{enumerate}
\item to construct the background,
\item to show that the $\bmPsi\neq0$ solution has a lower grand canonical potential, and
\item to derive the London equation. (This establishes that the $\bmPsi\neq0$ phase is a superconducting phase.)
\end{enumerate}

\subsection{Background solution}\label{sec:bg_low}

The solution \eqref{eq:analytic_sol} is the solution only at the critical point, and we first construct the background solution in the low-temperature phase. As mentioned in \sect{critical}, the construction is essentially the same as the $\zl=1$ case \cite{Herzog:2010vz}.

Consider the solution of the form 
\begin{align}
\bmPsi=\bmPsi(s)~, \quad\bmA_t=\bmA_t(s)~, \quad\bmA_i=\bmA_s=0~.
%
\end{align}
The equations of motion are given by
\begin{subequations}
\label{eq:eom_homo}
\begin{align}
&\del_s \left(\frac{h}{s}\del_s \bbPsi \right) + \left\{ \frac{\bbA_t^2}{hs^2} 
+ \frac{1}{s^3} \right\}\bbPsi =0~, \\
&\del_s^2 \bbA_t = \frac{2}{hs^2} \bbPsi^2\bbA_t~, \\
&\bmPsi^\dagger \bmPsi' - \bmPsi^{\dagger'} \bmPsi = 0~.
%
\end{align}
\end{subequations}
One can set $\bmPsi$ to be real.
We construct the background perturbatively:
\begin{subequations}
\begin{align}
\bbPsi &= \epsilon^{1/2} \left( \bmPsi_1 + \epsilon\, \bmPsi_2 + \cdots \right)~,
\label{eq:background_low_psi} 
\\
\bbA_t &= \bmPhi_0 + \epsilon \bmPhi_1 + \epsilon^2 \bmPhi_2 +\cdots~,
%
\end{align}
\end{subequations}
where $\epsilon$ is a small parameter whose meaning will be clear in a moment. From \sect{critical}, we already know
\begin{subequations}
\begin{align}
\bmPhi_0 &= 1-s~, \\
\bmPsi_1 &= \frac{s}{1+s}~.
%
\end{align}
\end{subequations}
To proceed to higher orders in $\epsilon$, we impose the boundary conditions following Ref.~\cite{Herzog:2010vz}:
\begin{itemize}
\item
$\bmPsi_n$: Asymptotically, no slow falloff and no fast falloff, or $\Psi_n^{(0)}=\Psi_n^{(1)}=0$ (for $n\geq2$). The former means the condition for a spontaneous condensate. The latter means that $\calO$ comes only from $\bmPsi_1$. At the horizon, we impose the regularity condition.
\item
$\bmPhi_n$: $\bmPhi_n(s=1)=0$ at the horizon.
\end{itemize}
Namely, we fix the fast falloff $\calO$, but the chemical potential is corrected as 
\begin{align}
\barmu = 1+\epsilon\delta\barmu_1+\epsilon^2\delta\barmu_2+\cdots~.
%
\end{align}
Under these boundary conditions, 
\be
\bbPsi\sim\epsilon^{1/2}s~,
%
\ee
so $\epsilon^{1/2}$ represents the order parameter $\calO$. We impose 4 boundary conditions in total, which completely fixes the solution. For example, $\bmPhi_1$ and $\bmPsi_2$ have 4 integration constants, and they are determined by the 4 conditions. 

At $O(\epsilon)$,
\begin{align}
\bmPhi_1
&=\delta\bar{\mu}_1(1-s)-\frac{s(1-s)}{2(1+s)} \\
&\sim \delta\barmu_1+(-1/2-\delta\barmu_1)s+\cdots~,
%
\end{align}
where we imposed the boundary condition $\bmPhi_1(s=1)=0$, and $\delta\barmu_1$ is the remaining integration constant. It is fixed at $O(\epsilon^{3/2})$ from the condition that $\Psi_2^{(0)}=0$.

At $O(\epsilon^{3/2})$, there are 2 more integration constants and $\delta\barmu_1$. After imposing the boundary condition at the horizon and the $\Psi_2^{(1)}=0$ condition, one obtains
\begin{align}
\bmPsi_2 &= -\frac{s^2}{3(1+s)^2} 
\\ & \quad%
+ \left(\delta\barmu_1-\frac{1}{12}\right) \frac{s\ln s}{2(1+s)} + \left(\frac{1}{4}-\delta\barmu_1\right)\frac{s\ln(1+s)}{1+s} 
\label{eq:psi2}
\nonumber \\ %
&\sim \frac{1}{2}\left(\delta\barmu_1-\frac{1}{12}\right) s\ln s~,
\end{align}
so the remaining no slow falloff condition $\Psi_2^{(0)}=0$ gives $\delta\bar{\mu}_1=1/12$. 
Then, at $O(\epsilon)$, the chemical potential becomes
\begin{align}
\bar{\mu} &= \bbA_t|_{s=0} 
\nonumber\\
&= 1+\frac{1}{12}\epsilon+\cdots~,
%
\end{align}
so $\epsilon$ is determined as 
\begin{align}
\epsilon=12\bepsmu = 12\epsmu/(2\zl)~.
%
\end{align}
Thus, 
\begin{align}
\calO = -(2\zl) \epsilon^{1/2} = -(24\zl\epsmu)^{1/2}~,
%
\end{align}
and the critical exponent $\beta=1/2$. More generally, $A_t^{(0)}$ gives the GL equation of motion (\sect{low_chi} and \sect{GL}).

At $O(\epsilon^{2})$,
\begin{align}
\bmPhi_2
&= - \frac{(1-s)}{1728(1+s)^2}(253+842s+253s^2) 
\nonumber \\ & \quad%
+ \frac{1}{36}(7-13s)\ln2 + \frac{\ln(1+s)}{3(1+s)}\\
&\sim \frac{-253+336\ln2}{1728} + \frac{493-624\ln2}{1728} s + \cdots \\
&=\delta\barmu_2 + \left(\frac{373}{864}-\frac{5}{9}\ln2+\delta\barmu_2 \right)s + \cdots~, \\
\delta\barmu_2&=\frac{-253+336\ln2}{1728}~.
%
\end{align}
Again, we determine an integration constant $\delta\barmu_2$ at $O(\epsilon^{5/2})$ from the condition $\Psi_3^{(0)}=0$. The expression for $\bmPsi_3$ is too cumbersome to write here.

\subsection{Grand canonical potential}\label{sec:grand_canonical}

We use the Lorentzian formalism to evaluate the grand canonical potential $\Omega$. (Note $S_\text{E}=\beta \Omega=-S_\text{L}$.) The matter on-shell action is given by
\begin{align}
S_\text{OS} =\int d^{p+1}x\, &\bigg\{ 
-\frac{p-\zl}{2} A_t^{(0)}A_t^{(1)} 
\nonumber \\ &
+ \int_0^1 du\, \sqrt{-g} g^{tt} \bmA_t^2 |\bmPsi|^2 
\bigg\}~.
%
\end{align}
We are interested in the grand canonical potential of the spontaneous condensate, or the solution with $\Psi^{(0)}=0$, so the boundary term from $\Psi$ vanishes. 

We evaluate the difference of the grand canonical potential between the $\bmPsi=0$ solution and the $\bmPsi\neq0$ solution. We fix the chemical potential as $\barmu=1+\epsilon\delta\barmu_1+\epsilon^2\delta\barmu_2+\cdots$, where $\delta\barmu_1$ and $\delta\barmu_2$ are obtained in the previous subsection. It turns out that  $\delta S_\text{OS}=0$ at $O(\epsilon)$, so we evaluate the difference at $O(\epsilon^2)$. This implies that one has to take into account up to $O(\epsilon^2)$ of $\bmA_t$ in order to evaluate the above boundary action. 

For the $\bmPsi=0$ solution,
\be
\bbA_t \sim (1+\epsilon \delta\barmu_1+\epsilon^2\delta\barmu_2+\cdots)(1-s)~.
%
\ee
In this case, only the boundary action contributes since $\bmPsi=0$. The on-shell action becomes
\begin{align}
S_{\Psi=0} &= \beta V_p (2\zl)^3 
\left\{ \frac{1}{2} + \delta\barmu_1\epsilon + \frac{1}{2}(\delta\barmu_1^2+2\delta\barmu_2)\epsilon^2 +\cdots \right\} \\
&= \beta V_p (2\zl)^3 \bigg\{ \frac{1}{2} + \frac{\epsilon}{12} 
\nonumber \\ & \quad %
+ \left( -\frac{247}{1728} + \frac{7}{36}\ln2 \right) \epsilon^2+\cdots \bigg\}~,
%
\end{align}
where $\beta$ is the inverse temperature, and $V_p$ is the boundary spatial volume. For the $\bmPsi\neq0$ solution,
\begin{align}
S_{\Psi\neq0} &= \beta V_p (2\zl)^3 
\bigg\{ \frac{1}{2} + \delta\barmu_1\epsilon 
\\ & \quad %
+ \left(-\frac{181}{1728} -\frac{1}{4}\delta\barmu_1+\frac{1}{2}\delta\barmu_1^2 + \frac{7}{36}\ln2 \right)\epsilon^2 +\cdots \bigg\} 
\nonumber \\ %
&= \beta V_p (2\zl)^3 \bigg\{ \frac{1}{2} + \frac{\epsilon}{12} 
\nonumber \\ & \quad %
+ \left( -\frac{211}{1728} + \frac{7}{36}\ln2 \right) \epsilon^2+\cdots \bigg\}~.
%
\end{align}
Thus, the difference is
\begin{align}
\delta S_\text{OS} &= S_{\Psi\neq0}-S_{\Psi=0}
\nonumber \\
&=\beta V_p \frac{\zl^3}{6}\epsilon^2 = -\beta \delta \Omega~,
\\
\Rightarrow\quad \frac{\delta\Omega}{V_p} &= -6\zl \epsmu^2~.
\label{eq:free_energy}
\end{align}
$\delta\Omega<0$, so the $\bmPsi\neq0$ solution is favorable. The difference is proportional to $\epsmu^2=(\mu-\mu_c)^2 \propto (\Tc-T)^2$, which implies the second-order phase transition. (The difference and its first derivative are continuous, but the second derivative is discontinuous.) The specific heat $C_\mu$ behaves as $C_\mu=-T\del^2\Omega/\del T^2 \propto T$, which determines the critical exponent $\alpha=0$, where $\alpha$ is defined by $C_\mu \propto (\Tc-T)^{-\alpha}$.

\subsection{Background with source}\label{sec:low_chi}

We construct the background without the source of the order parameter, but it is straightforward to extend the construction to the background with the source. Going back to \eq{psi2}, we obtained
\begin{align}
\bmPsi_2 &= -\frac{s^2}{3(1+s)^2} 
\\ & \quad%
+ \left(\delta\barmu_1-\frac{1}{12}\right) \frac{s\ln s}{2(1+s)} + \left(\frac{1}{4}-\delta\barmu_1\right)\frac{s\ln(1+s)}{1+s} 
\label{eq:psi2}
\nonumber \\ %
&\sim \frac{1}{2}\left(\delta\barmu_1-\frac{1}{12}\right) s\ln s~,
\end{align}
so the asymptotic behavior becomes
\be
\bbPsi \sim \frac{1}{2}\left(\delta\barmu_1-\frac{1}{12}\right) \epsilon^{3/2} s\ln s + \epsilon^{1/2} s~.
\label{eq:asymptotic_source}
\ee
Previously, we imposed the source-free condition $\Psi^{(0)}=0$, which gives $\delta\barmu_1=1/12$. We now allow $\Psi^{(0)} \neq 0$. The chemical potential is given by $\barmu = 1+\epsilon\delta\barmu_1$. At the critical point, $\barmu=1$, so $\delta\barmu_1=0$. From the asymptotic behavior \eqref{eq:asymptotic_source},  $\calO\propto\epsilon^{1/2}$ and $\Psi^{(0)} \propto \epsilon^{3/2}$. Then, the exponent $\delta$ defined by $\calO \propto (\Psi^{(0)})^{1/\delta} $ (at $\mu=\mu_c$) is $\delta=3$. 

One can evaluate the thermodynamic response function at low temperature. By imposing our boundary conditions, 
\be
\bar{\Psi}^{(0)} = \zl\left(\delta\barmu_1-\frac{1}{12}\right)\epsilon^{3/2}~.
%
\ee
The chemical potential is then determined as
\begin{align}
\barmu &= 1+\epsilon\delta\barmu_1 = 1 + \frac{1}{12}\epsilon + \frac{\bar{\Psi}^{(0)}}{\zl\epsilon^{1/2}}~,
\label{eq:chemical_source1}
\end{align}
which is rewritten as
\begin{align}
\bepsmu &= \frac{1}{12}\epsilon + \frac{\bar{\Psi}^{(0)}}{\zl \epsilon^{1/2}}~.
\label{eq:chemical_source}
\end{align}
This is essentially the GL equation of motion (\sect{GL}). For a fixed $\mu$, this gives
\be
d\epsilon = - \frac{12}{\zl \epsilon^{1/2}} d\bar{\Psi}^{(0)} + O(\bar{\Psi}^{(0)}d\epsilon)~.
%
\ee
Thus, 
\begin{align}
\chi_T^< &= \left.\frac{\del\calO}{\del\Psi^{(0)}} \right|_{\Psi^{(0)}=0}
\nonumber \\
&= \left.\frac{d\calO/d\epsilon}{d\Psi^{(0)}/d\epsilon} \right|_{\Psi^{(0)}=0}
=\frac{12}{2\zl\epsilon}
=\frac{1}{\epsmu}~,
\\
\Rightarrow A_< &=1~.
%
\end{align}
(Recall $\calO = -2\zl\epsilon^{1/2}$.) 

We obtained $\chi_T^<$ from the background solution, but it should also be possible to obtain it from the scalar perturbation as in \sect{high}. One would also obtain the full response function $\chi_{\omega,q}^<$ using the $(\epsilon,\omega,q)$-expansion. But, in the low-temperature phase, $\delta\Psi$ couple with $\delta A_t$ and $\delta A_x$, and the computation is more involved, so we leave it to a future work.

\subsection{Vector modes}\label{sec:vector}

From the vector mode, one can show the London equation and compute the conductivity. The $\delta A_y$-equation is given by
\begin{align}
&\del_s \left(\frac{h}{s^{1-1/\zl}}\del_s \delta A_y \right) 
\nonumber \\& %
+ \left\{ \frac{\bw^2}{hs^{2-1/\zl}} - \frac{\bq^2}{s^{3-2/\zl}} -\frac{2\bbPsi^2}{s^{3-1/\zl}} \right\}\delta A_y = 0~, 
%
\end{align}
where $\bbPsi$ was constructed in \eq{background_low_psi}. We impose the incoming-wave boundary condition at the horizon and $\delta A_y|_{s=0} = \sourceA{y}$ asymptotically. We again employ the ($\epsilon$, $\omega$)-expansion:
\begin{align}
\delta A_y &= (1-s)^{-i\bw/2} 
\nonumber \\ &\quad \times %
((a_c + \epsilon a_\epsilon +\cdots) + \bw a_\omega +\cdots)~.
%
\end{align}
The equation of motion reduces to
\begin{subequations}
\begin{align}
\calL_a a_c &= 0~, \\
\calL_a a_i &= \nj_i (a_c)~,
%
\end{align}
where
\begin{align}
\calL_a &= \del_s \left(\frac{h}{s^{1-1/\zl}}\del_s \right)~.
\end{align}
\end{subequations}
The homogeneous equation $\calL_a a_c =0$ can be solved as
\begin{align}
a_c &= c_1 
\quad \\ & %
+ c_2 s^{2-1/\zl} \frac{2\zl}{2\zl-1} 
{}_2F_1\left( 1,\frac{2\zl-1}{2\zl},\frac{4\zl-1}{2\zl};s^2 \right)~.
\nonumber
%
\end{align}
From the regularity at the horizon, $c_2=0$.

The source terms of inhomogeneous equations then become
\begin{subequations}
\begin{align}
\nj_\epsilon &= \frac{2}{s^{1-1/\zl}(1+s)^2} c_1~, \\
\nj_\omega &= -\frac{i}{2} \left( \frac{1+s}{s^{1-1/\zl}} \right)' c_1~.
%
\end{align}
\end{subequations}
Again, we discuss the $\zl=1$ and $\zl\neq1$ cases separately.

\subsubsection{$\zl=1$}
The solution is 
\begin{subequations}
\begin{align}
a_\epsilon &= c_1 \frac{1}{1+s} 
\sim c_1  (1-s)~, \\
a_\omega &=\frac{1}{2} i c_1\ln(1+s) 
\sim ic_1 s/2~.
%
\end{align}
\end{subequations}
The asymptotic behavior then becomes
\begin{align}
\delta A_y &= \frac{\sourceA{y}(k)}{1+\epsilon+\cdots} (1-s)^{-i\bw/2} 
\nonumber \\ & \quad\times %
\left\{ 1+ \frac{\epsilon}{1+s} + \frac{1}{2} i\bw \ln(1+s) + \cdots \right\}\\
&\sim \sourceA{y}\{ 1+(-\epsilon+ i\bw+\cdots) s \}~.
%
\end{align}
We determine the constant $c_1$ from the asymptotic boundary condition $\delta A_y|_{s=0} = \sourceA{y}$. So, 
\begin{align}
 J^y &= (4\zl-2) \vevA{y} \\
&= 2 (-\epsilon+ i\bw +\cdots) \sourceA{y}~.
%
\end{align}
The $\omega \to0$ limit gives the London equation 
\begin{align}
 J^y  = -(1/\lambda^2)\sourceA{y}~,
%
\end{align}
with the London penetration depth $\lambda^{-2}= 2\epsilon$. The conductivity is then given by
\begin{align}
\sigma(\omega) &= \frac{ J^y }{i\omega \sourceA{y}} 
\nonumber \\&
= \frac{2i\epsilon}{\omega}+ 1 +\cdots~.
%
\end{align}
$\text{Im}(\sigma)$ has the $1/\omega$-pole which implies the diverging DC conductivity.

A \SC\ has singular behaviors in the current, but its essence is not in the diverging DC conductivity but in the London equation. A diverging DC conductivity also appears in a perfect conductor, but the London equation is unique to \SCs.

When one combines the London equation with the Maxwell equation, one obtains the Meissner effect. However, for usual holographic superconductors, the boundary Maxwell field is added just as an external source and is not dynamical in the boundary theory, so the Meissner effect does not arise; the magnetic field can always penetrate into the material. In this sense, a \HSC may be regarded as a superfluid. (In low spatial dimensions $p\leq2$, one can obtain a boundary theory with a dynamical Maxwell field. See, \eg, Ref.~\cite{Domenech:2010nf}.)

However, the London equation must hold even in this case if the system is really a \SC\ or a superfluid. The London equation is the response of the current under the external source, and whether the source is dynamical or not is irrelevant to the issue.

\subsubsection{$\zl>1$}

For $a_\omega$, 
one can get the generic expression%
\footnote{We set $p=3\zl$, but the $O(\omega,q^2)$-equations can actually be solved for a generic $(p,\zl)$.}:
\begin{align}
a_\omega &=\frac{1}{2}ic_1
\bigg\{ s^{2-1/\zl} \frac{2\zl}{2\zl-1} {}_2F_1\left( 1,\frac{2\zl-1}{2\zl},\frac{4\zl-1}{2\zl};s^2 \right) 
\nonumber \\ & \quad %
+ \ln(1-s) \bigg\} \\
&\sim O(s) + ic_1 \frac{\zl}{2\zl-1} s^{2-1/\zl}~.
%
\end{align}
For $a_\epsilon$, the generic expression is either difficult to obtain or too cumbersome, but again the fast falloff has a simple expression:
\begin{align}
a_\epsilon^{(1)} &= -\frac{2\zl}{2\zl-1} c_1 I(\zl)~.
%
\end{align}
The asymptotic behavior then becomes
\begin{align}
\delta A_y &\sim A_y^{(0)} \big[ 1+ \cdots + 
\\ & \quad %
\frac{1}{4\zl-2}\left\{ -4\zl I(\zl) \epsilon+ i\omega +\cdots \right\} s^{2-1/\zl} + \cdots \big]~.
\nonumber %
%
\end{align}
Thus,
\begin{align}
 J^y  &= \left\{ -4\zl I(\zl) \epsilon+ i\omega + \cdots \right\}  \sourceA{y}~.
\label{eq:London}
\end{align}
Again, the $\omega\to0$ limit gives the London equation $ J^y  = -(1/\lambda^2) \sourceA{y}$ with the London penetration depth $\lambda^{-2} = 4\zl I(\zl) \epsilon$. 
The conductivity is then given by
\begin{align}
\sigma &= \frac{ J^y }{i\omega\sourceA{y}} 
\nonumber \\&
=  \frac{i}{\omega}\frac{I(\zl)}{\zl} \calO^2 + 1 +\cdots~.
%
\end{align}
For large $\zl$, 
\begin{align}
\sigma \approx \frac{i}{\omega}\calO^2 + 1 +\cdots~.
%
\end{align}

The GL parameter $\kappa$ is defined by
\be
\kappa^2 :=\left( \frac{\lambda}{\xi_>} \right)^2 = \frac{\zl}{24I(\zl)^2}~.
%
\ee
In conventional \SCs, $\kappa^2<1/2$ for type I and $\kappa^2>1/2$ for type II \SCs. For $\zl=1$, $\kappa^2=1/6$, so one may conclude that our system is type I (in the sense of $\kappa$), but whether our system is type I or II is more subtle. Physically, $1/\lambda$ represents the Maxwell field mass, so we should 
determine the normalization of $\lambda$ by comparing with normalization of the boundary Maxwell action. 
However, as mentioned above, the boundary Maxwell field is added as an external source here and is not dynamical in the boundary theory, so the normalization cannot be determined%
\footnote{The value of $\kappa$ for \HSCs has been discussed in Refs.~\cite{Maeda:2008ir,Dias:2013bwa}. In Ref.~\cite{Dias:2013bwa}, $\kappa$ depends on the scalar charge $e$. On the contrary, if we restore dimensionful parameters, our $\kappa$ does not depend on $e$ (\appen{restore}). }. 
(Holographic superconductors are type II \SCs\ in the sense that there is no Meissner effect.)

\section{Background with vector potential}\label{sec:bg_vector}

In this section, we add a vector potential $\bmA_i$ as a background. 
%
%
We again consider the perturbative expansion:
\begin{subequations}
\begin{align}
\bbPsi &= \epsilon^{1/2} \left(\bmPsi_1' + \cdots \right)~,
\\
\bbA_t &= \bmPhi_0'  + \cdots~, \\
\bbA_y &= \bmA_{y,0} + \epsilon \bmA_{y,1} +\cdots~.
%
\end{align}
\end{subequations}
Note that we take into account (1) $A_y$ as a background and (2) the backreaction of $\bmA_{y}$ onto the other fields. (That is why we use variables with primes.) The former is the difference from the perturbative expansion in \sect{bg_low}, and the latter is difference from the vector mode computation in \sect{vector}.

At $O(\epsilon^0)$, the Maxwell equation becomes 
\begin{align}
\nabla_N\bmF^{MN}_{~~~~~, 0}=0~,
%
\end{align}
where $\bmF_{MN,0} =\del_M \bmA_{N,0} - \del_N \bmA_{M,0}$. The equation has simple solutions. Namely, $\bmPhi_0'=\bmu(1-s)$ and two interesting solutions for $\bmA_{y,0} $:
\begin{subequations}
\begin{align}
\bmA_{y,0} &=\bay =\text{(constant)}~,   \\ 
\bmA_{y,0} &= \bar{B}x~.   
%
\end{align}
\end{subequations}
The former corresponds to adding a constant superfluid flow $a_y$, and the latter corresponds to adding a constant magnetic field $B$.

\subsection{Superfluid flow}

For the superfluid flow, it is enough to consider homogeneous perturbations. The equations of motion are given by
\begin{subequations}
\begin{align}
&\del_s \left(\frac{h}{s}\del_s \bbPsi \right) + \left\{ \frac{\bbA_t^2}{hs^2} - \frac{\bbA_y^2}{s^{3-1/\zl}} + \frac{1}{s^3} \right\}\bbPsi =0~, \\
&\del_s^2 \bbA_t = \frac{2}{hs^2} \bbPsi^2\bbA_t~, \\
&\del_s \left(\frac{h}{s^{1-1/\zl}}\del_s  \bbA_y \right) - \frac{2\bbPsi^2}{s^{3-1/\zl}}\bbA_y =0~.
%
\end{align}
\end{subequations}
We impose the same boundary conditions as \sect{bg_low}.

Our main interest is the phase diagram, \ie, the deviation of the critical point by the vector potential. Then, we evaluate how $\bmA_y$ at $O(\epsilon^0)$ affects $\bmPsi_1'$ at $O(\epsilon^{1/2})$. This in turn affects the value of $\mu_c$. We employ the $a_y$-expansion as well as the $\epsilon$-expansion \cite{Herzog:2010vz}. Namely, 
\begin{subequations}
\begin{align}
\bmPsi_1' &= \bmPsi_1 + \bay^2 \bmPsi_a + \cdots~, \\
\bm\Phi_0' &= (1 + \bay^2c_0+\cdots)(1-s)~,
%
\end{align}
\end{subequations}
where $c_0$ is a constant. This expansion is consistent with the above equations of motion.

At $O(\epsilon^{1/2})$, $\bmPsi_1=s/(1+s)$. The $\bmPsi_a$-equation becomes
\begin{subequations}
\begin{align}
\calL_\psi \bmPsi_a &= \nj_a~, \\
\nj_a &= -2c_0 \frac{1-s}{s(1+s)^2}+\frac{1}{s^{2-1/\zl}(1+s)}~.
%
\end{align}
\end{subequations}
The equation is hard to solve in general. However, to determine the $\bay$-dependence on the chemical potential, it is enough to obtain the slow falloff of $\bmPsi_a$. 
The slow falloff has a simple expression:
\begin{align}
\bmPsi^{(0)}_a &= -\int_0^1 ds\, \nj_a\frac{s}{1+s} 
\\
&= \frac{c_0}{2} - I(\zl)~.
%
\end{align}
We impose the boundary condition $\bmPsi^{(0)}_a=0$, which gives $c_0=2I(\zl)$. 

At the critical point, the order parameter vanishes, so $\epsilon=0$. Then, to determine the critical point, set $\epsilon=0$, and $\sourceA{t}$ gives the critical chemical potential:
\begin{align}
\mu_c &= \mu_{c,0}+a_y^2 \frac{I(\zl)}{\zl}+\cdots 
\label{eq:flow} \\
&\sim \mu_{c,0} + a_y^2 \quad (\zl\gg1)~,
\end{align}
where $\mu_{c,0}=2\zl$ is the critical point without superfluid flow.

To obtain $J^\mu$, one needs to obtain $\bmA_\mu$. This is necessary to derive the second sound $c_2$ \cite{Herzog:2008he,Herzog:2010vz}:
\be
c_2^2 = - \left.\frac{\del J^y/\del a_y}{\del J^t/\del\mu}\right|_{a_y=0}~.
%
\ee
To derive $c_2$, it is enough to use the results obtained in the previous section. The $\bmA_{y,1}$-equation is the same as the vector mode perturbation $a_\epsilon$. So, $J^y$ is given by
\begin{align}
\bbA_y &\sim \bay \left[ 1 - 
\frac{1}{4\zl-2} 4\zl I(\zl) \epsilon s^{2-1/\zl} + \cdots \right]~,
\nonumber \\
\Rightarrow J^y  &= (4\zl-2)\vevA{y} = -4\zl I(\zl) \epsilon\, a_y~.
%
\end{align}
$J^t$ is given by
\begin{align}
\bbA_t &\sim (1-s) + \epsilon\{ \delta\barmu_1+(-1/2-\delta\barmu_1)s \} +\cdots 
\nonumber \\
&=\bmu+(6-7\bmu)s +\cdots~, 
\nonumber \\
\Rightarrow J^t &= -2\zl\vevA{t} = 2\zl(7\mu-12\zl)~.
%
\end{align}
Thus,
\begin{align}
c_2^2 
&= \frac{I(\zl)}{14\zl^2}\calO^2 \\
&\sim \frac{1}{14\zl}\calO^2 \quad (\zl \gg1)~.
%
\end{align}


%
%

%
%
%
%

\subsection{Magnetic field}

We follow Ref.~\cite{Maeda:2009vf} to obtain the critical magnetic field. The $\bmPsi$-equation is given by
\begin{align}
\del_s \left(\frac{h}{s}\del_s \bbPsi \right) + \left\{ \frac{\bbA_t^2}{hs^2} + \frac{1}{s^{3-1/\zl}}(\bdel_i-i\bbA_i)^2 + \frac{1}{s^3} \right\}\bbPsi =0~.
%
\end{align}
Here, $\bdel_i =\del_i/(2\zl) = \del/\del\bx^i$, where $\bx^{i} :=(2\zl) x^i$.
The vector potential is given by $\bbA_y=\bar{B}x=\bbB \bx$, where $\bbB:=B/(2\zl)^2$.

This problem can be solved as a Landau-level problem after separation of variables. First, set%
\footnote{For simplicity, we set the other momenta as $k_3=k_4=\cdots=0$.}
\[
\bbPsi(x,y,s)=e^{i k_y y}\varphi(x,s;k_y)~.
%
\]
The equation then takes the form
\begin{align}
s^{3-1/\zl} & \left[ \del_s \left(\frac{h}{s} \del_s \right) + \frac{\bbA_t^2}{hs^2} + \frac{1}{s^3} \right]\varphi 
\nonumber \\ & \quad %
= \left[ -\del_{\bx}^2+\left(\bky-\bbB\bx\right)^2 \right] \varphi~,
%
\end{align}
so setting 
\[
\varphi_n(x,s;k_y)=\rho_n(s)\gamma_n(x;k_y)~,
%
\]
one obtains
\begin{subequations}
\begin{align}
&\left( -\del_X^2+X^2 \right) \gamma_n = \lambda_n\gamma_n~, \\
& \left[ \del_s \left(\frac{h}{s} \del_s \right) + \frac{\bbA_t^2}{hs^2} + \frac{1}{s^3} \right] \rho_n = \bbB \lambda_n \frac{\rho_n}{s^{3-1/\zl}}~,
%
\end{align}
\end{subequations}
where $X:= \sqrt{\bbB}(\bx-\bky/\bbB)$, and $\lambda_n$ is a separation constant. The $\gamma_n$-equation is solved by the Hermite function $H_n$ as
\be
\gamma_n(X)= e^{-X^2/2} H_n(X)~,
%
\ee
with eigenvalue $\lambda_n$ as
\begin{align}
\lambda_n=2n+1 \quad (n\geq0)~.
%
\end{align}
The solution is parametrized by $\bbB\lambda_n$, so one has the largest magnetic field $B_{c2}$ when $\lambda_n$ takes the minimal value, namely the $n=0$ solution.

The $\rho_0$-equation is given by
\begin{align}
\del_s \left(\frac{h}{s}\del_s \rho_0 \right) + \left\{ \frac{\bbA_t^2}{hs^2} -\frac{\bbB_{c2}}{s^{3-1/\zl}} + \frac{1}{s^3} \right\}\rho_0 =0~.
%
\end{align}
Then, the problem formally reduces to the same problem as the superfluid flow one with the replacement $\bay^2$ by $\bbB_{c2}$. Thus, the critical point is given by
\be
\mu_c=\mu_{c,0}+B_{c2}\frac{I(\zl)}{\zl}~,
%
\ee
Using the result of $\xi_>^2$ in \sect{high}, we get
\be
B_{c2}=1/\xi_{>}^2~.
\label{eq:Bc2}
\ee

\section{The dual GL theory}

\subsection{Identifying the dual GL theory}\label{sec:GL}

We thus obtained all critical exponents and critical amplitudes
\begin{subequations}
\begin{align}
(\alpha,\beta,\gamma,\delta,\nu,\eta,\zd) &= \left(0,\frac{1}{2},1,3,\frac{1}{2},0,2\right)~, 
\\ 
A_>&=2A_<~.
%
\end{align}
\end{subequations}
The results are consistent with the standard GL theory or the $\phi^4$ mean-field theory. 
In fact, the following GL theory reproduces all our results%
\footnote{As always, presumably the dual theory is some large-$N_c$ gauge theory. This is the effective GL theory at low energy and momentum.}:
\begin{subequations}
\begin{align}
H_\text{GL} &= \int d^px\, \bigg\{ \frac{\cK}{2} |D_i\phi|^2 
-\frac{\ctwo}{2}\epsmu\,|\phi|^2 + \frac{\cfour}{4} |\phi|^4 +\cdots 
\nonumber \\ & \quad %
- \cphi (\phi J^\dag+\phi^\dag J)\bigg\}~, 
\\
D_i &:= \del_i -i\cA \sourceA{i}~.
%
\end{align}
\end{subequations}
In the dynamic case, consider the time-dependent GL equation (for Model A dynamic universality class):
\begin{align}
\Gamma^{-1} \del_t\phi &= - \frac{\delta H_\text{GL}}{\delta\phi} \\
&= -\frac{\cK}{2}D_i^2\phi - \frac{\ctwo}{2}\epsmu\phi + \frac{\cfour}{2} \phi|\phi|^2 - \cphi J~.
%
\end{align}
We determine the GL parameters $(\ctwo,\cfour,\cphi,\cK,\cA,\Gamma)$ to reproduce our holographic results. 

In the static homogeneous case, the $\phi$-equation becomes
\be
\ctwo \epsmu\phi - \cfour \phi|\phi|^2 + 2\cphi J=0~.
\label{eq:EOM_GL}
\ee
Substituting the $J=0$ solution $|\phi|^2=\ctwo\epsmu/\cfour$ into $H_\text{GL}$, one obtains the grand canonical potential:
\be
\frac{\Omega}{V} = - \frac{\ctwo^2}{4\cfour}\epsmu^2~.
\label{eq:free_energy_GL}
\ee
The current is given by
\be
J^i := -\frac{\delta H_\text{GL}}{\delta \sourceA{i}} = -\cK\cA^2|\phi|^2 \sourceA{i}~.
%
\ee
In the high-temperature phase, the response function is given by
\be
\chi_k^> = \frac{ \cphi \Gamma }{ -i\omega+\frac{\Gamma}{2}(\cK q^2+\ctwo|\epsmu|)}~,
%
\ee
which implies
\begin{align}
\omega &= -i\frac{\Gamma}{2}\ctwo (|\epsmu|+\frac{\cK}{\ctwo}q^2)~, \\
\xi_>^2 &=\frac{\cK}{\ctwo} |\epsmu|^{-1}~.
%
\end{align}
Add a background vector potential. When a constant $A_y=a_y$ is added, the critical point is shifted as
\be
\mu_c =\mu_{c,0} + \frac{\cK}{\ctwo} \cA^2 a_y^2~.
\label{eq:flow_GL}
\ee
When a magnetic field is added, the critical magnetic field is given by
\be
B_{c2} = \frac{\ctwo}{\cK\cA} \epsmu = \frac{1}{\cA} \xi_>^{-2}~,
\label{eq:Bc2_GL}
\ee
by solving the Landau-level problem.

Returning to our holographic results, \eq{chemical_source} is rewritten as 
\begin{align}
& \epsmu \calO - \frac{1}{24\zl} \calO^3 + 2 \Psi^{(0)} = 0~,
\label{eq:EOM_holo}
\end{align}
which takes the form of the GL theory equation of motion \eqref{eq:EOM_GL}. The grand canonical potential, the current, and the response function are obtained in Eqs.~\eqref{eq:free_energy}, \eqref{eq:London}, and \eqref{eq:response_holo}, respectively. These determine the GL parameters as
\begin{subequations}
\begin{align}
\cal{H} &= \frac{I(\zl)}{2\zl} |(\del_i-i\sourceA{i})\phi|^2  
- \frac{\epsmu}{2} |\phi|^2
+ \frac{1}{96\zl} |\phi|^4 +\cdots
\nonumber \\ & \quad %
- (\phi J^\dag+\phi^\dag J)~, \\
\frac{\Gamma}{2} &= \frac{1+3i}{5}~.
%
\end{align}
\end{subequations}
%
%
In the presence of a background vector potential, Eqs.~\eqref{eq:flow} and \eqref{eq:Bc2} agree with Eqs.~\eqref{eq:flow_GL} and \eqref{eq:Bc2_GL}, respectively.  

One would be tempted to ask how various results change as we vary $\zl$. But to make such a comparison, one must keep in mind that (1) we consider a special class of Lifshitz theories and (2) we must specify what quantities to fix as we vary $\zl$.

We consider a special class of theories where $p=3\zl$ and $m^2=-4z^2$. Even the spatial dimensionality $p$ is different for a different $\zl$, and it is unclear if the comparison with a different $\zl$ is physically meaningful. Also, some results may be generic for Lifshitz theories in general but some are not. As a simple example, in our case, $(\mu/T)_c$ is independent of $\zl$. This is so by construction of our theories as discussed in \sect{critical} and is certainly not a generic behavior. It simply means that \LHSCs\ have enough parameters to fix  $(\mu/T)_c$ as we vary $\zl$.

We also have to specify what quantities to fix. One natural candidate is $\mu_c$ (or $T_c$) since $(\mu/T)_c$ is $\zl$-independent, but it is unclear if this is really appropriate. We do not have the answer to this question. So far we set $\rh=1$ just for simplicity, so here we simply fix $\rh$ (and $\epsmu$) and how various results change as we vary $\zl$. Again, we do not mean that fixing $\rh$ is natural from the boundary point of view. Rather, the following comparison should be regarded as a handy way to understand the $\zl$-dependence of our holographic results or the dual GL theory.
\begin{enumerate}
\item
In the dual GL theory, the coefficient of the $\phi^4$-term becomes smaller as we increase $\zl$. So, the condensate increases as $|\phi|^2 \propto \zl$. 
\item
The $\zl$-dependence appears only in the kinetic term and the $\phi^4$-term. Thus, the relaxation time $\tau_{q=0}$ of the order parameter does not depend on $\zl$. 
\item 
The correlation length $\xi$ depends on the kinetic term so does depend on $\zl$. It monotonically increases as $I(\zl)/\zl$ but increases slowly and reaches a constant value for $\zl\gg1$. 
\item
On the other hand, the London penetration depth $\lambda$ decreases. This is because $J^i=-(1/\lambda^2) \sourceA{i} \propto -|\phi|^2 \sourceA{i}$ and because $\phi$ increases. ($\lambda$ also depends on the kinetic term so has the factor $I(\zl)/\zl$, but it is not a dominant factor.) Then, the GL parameter $\kappa$ decreases.
\item
As usual, the presence of a background vector potential $\sourceA{i}$ increases the critical chemical potential $\mu_c$. From the gravity point of view, this is because $A_i$ increases the effective mass of $\Psi$. $\mu_c$ monotonically increases as $I(\zl)/\zl$ since $\sourceA{i}$ comes from the covariant derivative in the kinetic term.
\end{enumerate}

\subsection{Lifshitz exponent and dynamic critical exponent}\label{sec:two_z}

We already mentioned that some results are not generic to Lifshitz theories in general. Then, what results are expected to be generic? An obvious answer is critical exponents and the ratio of critical amplitudes. The $\phi^4$ mean-field critical exponents are likely to hold for theories of \eq{Lifshitz_HSC}. The dynamic critical exponent $\zd=2$ is also likely to hold.

For a Lifshitz geometry, one would expect a dispersion relation of the form
\be
\omega \propto q^{\zl}~.
%
\ee
This form is expected from the Lifshitz scaling \eqref{eq:scaling_Lifshitz}. But from the analysis of the high-temperature phase, the order parameter obeys the dispersion relation 
\be
\omega \propto q^2~,
%
\ee
\ie, the dynamic critical exponent $\zd=2$ irrespective of the value of the Lifshitz exponent $\zl$. This does not contradict with the Lifshitz scaling. If we restore the horizon radius $\rh$,
\be
\omega \propto \rh^{\zl-2}q^2~.
%
\ee
Namely, at finite temperature, there are two length scales $\rh$ and $1/q$. They combine to give the scaling dimension $\zl$. In other words, the Lifshitz scaling alone does not determine the dynamic critical exponent. 

Then, what determines $\zd$? We obtain $\zd=2$ because the order parameter is not a conserved charge. According to the classification of Hohenberg and Halperin \cite{Hohenberg:1977ym}, all our models belong to Model A universality class.

The dynamic universality class is classified based on
\begin{enumerate}
\item whether the order parameter is conserved or not,
\item whether there are the other hydrodynamic modes which couple to the order parameter (none for Model A and B).
\end{enumerate}
Conservation laws play important roles to determine the dynamic universality class since a conservation law forces the relaxation to proceed more slowly. When only the order parameter matters in critical dynamics, a nonconserved order parameter gives Model A, and a conserved order parameter gives Model B.

The Lifshitz geometry is conjectured to describe a quantum critical point. Using \LHSCs, one prepares a new finite-temperature critical point in addition to the Lifshitz critical point. What we have shown is that the dynamic critical exponent $\zd$ associated with the new critical point can differ from $\zl$. Instead, the value of $\zd$ is determined by the critical dynamics of the new critical point.

\section{Discussion}

\subsection{Lifshitz geometry and \HSCs}

The Lifshitz geometry appears even in the context of the standard $\zl=1$ \HSC \cite{Gubser:2009cg,Horowitz:2009ij}. Consider the backreaction of matter fields onto the geometry. In the high-temperature phase, $\Psi=0$, so the geometry is the Reissner-Nordstr\"{o}m AdS black hole. In the low-temperature phase, $\Psi\neq0$, but one may expect that the geometry is somewhat similar to the Reissner-Nordstr\"{o}m AdS black hole. However, the $T=0$ geometry is conjectured to be a Lifshitz geometry in the IR and the AdS geometry in the UV. The solution in IR has been constructed, but the full geometry remains an open question.

It is unclear what happens at low temperature, but it is natural to expect that a Lifshitz-like \bh appears at low enough temperature. (Unfortunately, the Lifshitz \bh used in this paper is not a solution of the Einstein-Maxwell-complex scalar system.) Then, one should consider the Einstein-Maxwell-complex scalar system in a Lifshitz (IR)-AdS (UV) black hole. This is not an easy task however. First, the full geometry is not constructed even in the $T=0$ limit. Second, the stability of the geometry is an different issue. Finally, one has to solve perturbations in the full geometry to explore various properties. 

What we have done in this paper is one small step towards this program; we solved matter fields in a simple Lifshitz \bh background. As we have seen in this paper, qualitative behaviors of \LHSCs\ are the same as the ones of the standard holographic superconductors. In particular, static and dynamic critical exponents are the same. One would expect those behaviors are common even in the full problem. Critical dynamics is governed by dynamics of the critical point itself (such as criteria 1 and 2 in the previous subsection) and is not governed by the Lifshitz exponent $\zl$ in the underlying geometry. 

\subsection{Implications to quantum criticality}

We briefly discuss the implications of our result on quantum criticality. The Lifshitz geometry is conjectured to describe a quantum critical point. In this sense, our system has two critical points:
\begin{itemize}
\item
One is the $T=0$ quantum critical point. Its dynamic scaling is determined by $\zl$.
\item
The other is the $T\neq0$ superconducting critical point explored in this paper. Its dynamic scaling is determined by $\zd$ as we have shown in this paper. 
\end{itemize}

It has been proposed that quantum criticality explains strange metallic behaviors of high-$T_c$ superconductors. According to the proposal, a quantum critical point is ``hidden" inside the superconducting dome, and the quantum criticality explains scaling behaviors of various transport coefficients even in the normal phase. 
 
Our model is far from real materials, but roughly speaking, the quantum critical point could correspond to the $T=0$ Lifshitz geometry, and the superconducting dome could correspond to the \LHSC. The Lifshitz scaling may determine the scaling behaviors in the normal phase. But our result implies that the Lifshitz scaling does not determine the scaling behavior of the order parameter near $T_c$. Rather, the $T\neq0$ critical point has its own scaling. Whatever the value of $\zl$ a quantum critical point has, the $T\neq0$ critical point is likely to have $\zd=2$ at the mean-field level.

\begin{acknowledgments}

M.\ N.\ would like to thank Joe Polchinski for his continuous support since M.\ N.\  was a graduate student. His inspiration and insight into physics were of inestimable importance to M.\ N.\ .
This research was supported in part by a Grant-in-Aid for Scientific Research (17K05427) from the Ministry of Education, Culture, Sports, Science and Technology, Japan. 

\end{acknowledgments}

\appendix

\section{Field/operator correspondence and holographic renormalization}\label{sec:dict}

The asymptotic behaviors of matter fields are given by
\begin{subequations}
\begin{align}
A_t & \sim A_t^{(0)}  + A_t^{(1)} \tilu^{p-\zl} 
\nonumber \\& %
\Rightarrow  A_t^{(0)}  + A_t^{(1)} \tils~, \\
A_i & \sim A_i^{(0)}  + A_i^{(1)} \tilu^{p+\zl-2} 
\nonumber \\& %
\Rightarrow A_i^{(0)}  + A_i^{(1)} \tils^{2-1/\zl}~, \\
\Psi &\sim \Psi^{(0)} \tilu^\Delta \ln u +\Psi^{(1)} \tilu^\Delta 
\nonumber \\& %
\Rightarrow \frac{\Psi^{(0)}}{2\zl} \tils\ln s +\Psi^{(1)} \tils~.
%
\end{align}
\end{subequations}
(In expressions after ``$\Rightarrow$", we set $p=3\zl=3\Delta/2$ and used $\tils:=\tilu^{2\zl}$.) 

The field/operator correspondence is derived by evaluating the on-shell action. The bulk on-shell action, in general, diverges, and one needs to add counterterm actions. We take the probe limit, so we discuss counterterm actions for matter fields only. We use the Lorentzian formalism. 

In the static homogeneous case, or at the leading order in the $(\omega,q)$-expansion, the scalar action diverges, and the counterterm action is
\begin{subequations}
\begin{align}
S_\text{CT} &= \frac{1}{e^2} \int_{\del\calM} d^{p+1}x\, L_1~, \\
L_1 &= -\frac{1}{L} \sqrt{-\gamma} \left(\Delta+\frac{1}{\ln\delta} \right)|\Psi|^2 
\nonumber \\ & %
\Rightarrow -\frac{2\zl}{L} \sqrt{-\gamma} \left(1+\frac{1}{\ln\deltas} \right)|\Psi|^2~,
%
\end{align}
\end{subequations}
where $\gamma_{\mu\nu}$ is the $(p+1)$-dimensional boundary metric and 
$u=\delta$ (or $s=\deltas:=\delta^{2\zl}$) is the UV cutoff. 
As usual, the second term is necessary for the scalar which saturates the BF bound. 
 
Using the standard holographic technique, one then gets
\begin{subequations}
\begin{align}
\rho &= -\frac{p-\zl}{e^2L} A_t^{(1)}
\Rightarrow -\frac{2\zl}{e^2L} A_t^{(1)}~,\\
 J^i &= \frac{p+\zl-2}{e^2L} A_i^{(1)}
\Rightarrow \frac{4\zl-2}{e^2L} A_i^{(1)}~,\\
\calO &= -\frac{1}{e^2L} \Psi^{(1)}
\Rightarrow -\frac{1}{e^2L} \Psi^{(1)}~.
%
\end{align}
\end{subequations}
(More precisely, left-hand sides represent expectation values such as $\bra\calO\ket$.)

The Lifshitz scaling \eqref{eq:scaling_Lifshitz} is just a coordinate transformation from the bulk point of view. The Maxwell field is a one-form, and $\Psi$ is a scalar, so they transform as
\be
A_t \to A_t/a^{\zl}~, A_i \to A_i/a~, \Psi \to \Psi~,
%
\ee
under the scaling. Then, the scaling dimensions are 
\begin{subequations}
\begin{align}
[\mu]_s &= \zl~, [\rho]_s = p~, \\
[A_i^{(0)}]_s &= 1~, [J^i]_s = p+\zl-1~, \\
[\Psi^{(0)}]_s &= [\calO]_s = \Delta~.
%
\end{align}
\end{subequations}
On the other hand, the mass dimensions are 
\begin{subequations}
\begin{align}
[\mu] &= [A_i^{(0)}] = [\Psi^{(0)}] = \text{M}~, 
\\ 
[\rho] &= [J^i] = [\calO] = \text{M}^p~.
%
\end{align}
\end{subequations}
(We choose mass dimensions as $[e^2] = \text{M}^{2-p}$ and $[A_M] = [\Psi] = \text{M}$.) 

Continuing higher orders in the derivative expansion, one has additional counterterms:
\begin{subequations}
\begin{align}
L_2 &= \frac{1}{4}M_F(\Psi) \sqrt{-\gamma}\,\gamma^{\mu\nu}\gamma^{\rho\sigma}F_{\mu\sigma}F_{\nu\sigma}~, \\
L_3 &= M_\Psi(\Psi) \sqrt{-\gamma}\,\gamma^{\mu\nu} (D_\mu\Psi)^\dag D_\nu\Psi~.
%
\end{align}
\end{subequations}
\noindent
$M_F$ and $M_\Psi$ are power series in $\Psi$ whose explicit forms are not necessary in the discussion below. 
In the text, we take into account $O(\omega,q^2)$ terms in the scalar perturbation and $O(\omega)$ term in the vector perturbation, so it is enough to consider $L_2$ and $L_3$, but they make no contribution.
For the scalar perturbation in the high-temperature phase, $F_{\mu\nu}=0$, so $L_2=0$, and
\begin{align}
L_3 &= \left(\frac{2\zl}{L}\right)^2 
M_\Psi h^{1/2} \left\{ - \frac{(\bw+\bbA_t)^2}{hs} + \frac{\bq^2}{s^{2-1/\zl}} \right\} |\delta\Psi(k)|^2 
\nonumber \\
&\sim O\left(s(\ln s)^2\right) + O\left(s^{1/\zl}(\ln s)^2\right)~,
%
\end{align}
so $L_3$ makes no contribution as $s\to0$ (for a finite $\zl$).
For the vector perturbation, $L_2 \sim O(\omega^2,q^2)$, so $L_2$ makes no contribution%
\footnote{Since $L_2$ is a relevant operator for $z\geq 1$ in our theories, it should be taken into account at $O(\omega^2, q^2)$.}, 
and 
\begin{align}
L_3 &= \left(\frac{2\zl}{L}\right)^2 M_\Psi \frac{h^{1/2}}{s^{2-1/\zl}} |\bbPsi|^2 \delta A_y(-k) \delta A_y(k) \nonumber \\
&\sim \epsilon\, O(s^{1/\zl})~,
%
\end{align}
where we consider the case of the spontaneous condensate for $\bbPsi$.

\section{Extracting falloffs}\label{sec:Green}

We solve the following differential equation:
\begin{subequations}
\begin{align}
\calL\varphi&=\nj~,
\label{eq:inhomo} \\
\calL &=\del_s(p(s)\del_s)~.
%
\end{align}
\end{subequations}
Denote two independent solutions of the homogeneous equation $\calL\varphi=0$ as $\varphi_1$ and $\varphi_2$. We assume that $\varphi_1$ satisfies the boundary condition at the horizon $s=1$. The solution of the inhomogeneous equation \eqref{eq:inhomo} which is regular at the horizon is given by
\begin{align}
\varphi (s) &=
 - \varphi_1(s) \int^s_0 ds'\, \frac{\nj(s') \varphi_2(s')}{p(s')W(s')}
 \nonumber \\ & \quad %
 - \varphi_2(s) \int^1_s ds'\, \frac{\nj(s') \varphi_1(s')}{p(s')W(s')}~,
\label{eq:inhomo_sol} 
%
\end{align}
where $W$ is the Wronskian
$W(s) := \varphi_1 \varphi_2' - \varphi_1' \varphi_2 $.

For example, for $\delta\Psi$ and $\delta A_y$,
\begin{align}
\delta\Psi: \varphi_1 &= \frac{s}{1+s}~, 
\nonumber\\
\varphi_2 &= \frac{s}{1+s} \ln\left[ \frac{s}{(1-s)^2}\right] \sim s \ln s~, 
\nonumber\\
p(s) &= \frac{h}{s}~. 
\nonumber\\
\delta A_y: \varphi_1 &= \frac{\zl}{2\zl-1}~,  
\nonumber\\
\varphi_2 &= s^{2-1/\zl} {}_2F_1\left( 1,\frac{2\zl-1}{2\zl},\frac{4\zl-1}{2\zl};s^2 \right) 
\nonumber\\
&\sim s^{2-1/\zl}~, 
\nonumber\\ 
p(s) &=\frac{h}{s^{1-1/\zl}}~.
\nonumber
%
\end{align}
For both cases, $pW=1$.

Even if the integral \eqref{eq:inhomo_sol} is difficult to evaluate or has a cumbersome expression, one can  extract a falloff. Suppose that $\varphi_2$ has the appropriate falloff. Then, near the AdS boundary $s\to\deltas$, 
\be
\varphi(\deltas) \sim - \varphi_2(\deltas) \int^1_{\deltas} ds\, \nj(s)\varphi_1(s)~.
\label{eq:inhomo_bdy}
\ee
This integral essentially gives the falloff coefficient we want. 

The $\deltas$-dependence in the integral essentially has no contribution from the following reason. First, the integral may or may not converge:
\begin{enumerate}
\item
When it converges, one can take the $\deltas\to0$ limit since the $\deltas$-dependence in the integral does not produce an appropriate falloff when it is combined with $\varphi_2(\deltas)$; it gives a subleading falloff.
\item
When it diverges, simply discard the $\deltas$-dependence in the integral since again it does not produce an appropriate falloff%
\footnote{There may be an exception. The $\deltas$-dependence in the integral may produce an appropriate falloff when it is combined with the \textit{subleading} term of $\varphi_2(\deltas)$.}. Even if it diverges as $\deltas\to0$, the expression \eqref{eq:inhomo_sol} itself does not.
\end{enumerate}

For example, the slow falloff of $\psi_q$ becomes
\begin{align}
\psi_q^{(0)} &= -\int^1_0 ds\, \nj_q \frac{s}{1+s} = -c_1 \int^1_0 \frac{ds}{s^{1-1/\zl}(1+s)^2}
\nonumber \\
&= - c_1 I(\zl)~.
%
\end{align}
Similarly, the fast falloff of $a_\epsilon$ becomes
\begin{align}
a_\epsilon^{(0)} &= - \int^1_0 ds\, \nj_\epsilon \frac{\zl}{2\zl-1} = - \frac{2\zl}{2\zl-1} c_1 \int^1_0 \frac{ds}{s^{1-1/\zl}(1+s)^2} 
\nonumber \\
&= - \frac{2\zl}{2\zl-1} c_1 I(\zl)~.
%
\end{align}

\section{Restoring units}\label{sec:restore}

We set $e=L=\rh=1$, but here we present some of our main results by restoring units. 
\begin{itemize}
\item
The scalar mode (high-temperature phase): the dispersion relation, the relaxation time, and the correlation length are given by
\begin{subequations}
\begin{align}
\omega &= \frac{3-i}{5}\left\{ |\epsmu|+\frac{I(\zl)}{\zl} L\left(\frac{\rh}{L}\right)^{\zl-2} q^2+\cdots \right\}~, \\
\tau^{-1} &= \frac{1}{5}|\epsmu| = \frac{I(\zl)}{5\zl} L\left(\frac{\rh}{L}\right)^{\zl-2} \xi_>^{-2}~, \\
\xi_>^2 &= \frac{I(\zl)}{\zl} L\left(\frac{\rh}{L}\right)^{\zl-2} \frac{1}{|\epsmu|}~.
%
\end{align}
\end{subequations}

\item
The order parameter:
\begin{subequations}
\begin{align}
\calO &= -\frac{1}{e^2L} \left(\frac{\rh}{L}\right)^{2\zl} \frac{2\zl}{L} \epsilon^{1/2} \\
&= - \frac{1}{e^2L} \left(\frac{\rh}{L}\right)^{3\zl/2} \left(\frac{24\zl}{L} \epsmu\right)^{1/2} \\
&= - \frac{\sqrt{48}\pi^2}{e^2} \Tc^{3/2}(\Tc-T)^{1/2} \quad (\zl=1)~.
%
\end{align}
\end{subequations}

\item
The current (low-temperature phase):
\begin{subequations}
\begin{align}
 J^y  &= \frac{1}{e^2L} \left(\frac{\rh}{L}\right)^{4\zl-2} 
\nonumber \\
&\times \bigg\{ -\frac{I(\zl)}{\zl} (4\zl^2\epsilon) 
+ \frac{i\omega L}{\left(\frac{\rh}{L}\right)^{\zl}} 
+\cdots \bigg\}  \sourceA{y} \\
&= -\frac{L}{e^2}  \left\{ \frac{1}{L}\left(\frac{\rh}{L}\right)^{3\zl-2} 24I(\zl) \epsmu \right\}  \sourceA{y} \quad (\omega=0)\\
&= -\frac{I(\zl)}{\zl} \frac{e^2L^3}{ \left(\frac{\rh}{L}\right)^2 } \calO^2 \sourceA{y}  \quad (\omega=0)~.
%
\end{align}
\end{subequations}
In our conventions, it is natural to define $\lambda$ as
\begin{align}
J^y &= -\frac{L}{e^2} \frac{1}{\lambda^2}\sourceA{y}~.
%
\end{align}
Then, $\lambda$ and $\kappa$ are given by
\begin{subequations}
\begin{align}
\lambda^2 &= L \left(\frac{\rh}{L}\right)^{-3\zl+2}\frac{1}{24I(\zl)\epsmu}~, \\
\kappa^2 &:= \left(\frac{\lambda}{\xi_>}\right)^2 
= \frac{\zl}{24I(\zl)^2}\left(\frac{\rh}{L}\right)^{-4\zl+4}~.
%
\end{align}
\end{subequations}

\item
The dual GL theory:
\begin{subequations}
\begin{align}
\cal{H} &= \frac{I(\zl)}{2\zl}L\frac{(eL)^2}{\left(\frac{\rh}{L}\right)^2} |(\del_i-i\sourceA{i})\phi|^2  
\nonumber \\ & \quad %
- \frac{\epsmu}{2} \frac{(eL)^2}{\left(\frac{\rh}{L}\right)^{\zl}} |\phi|^2
+ \frac{1}{96\zl} \frac{(eL)^6}{\left(\frac{\rh}{L}\right)^{4\zl}L}  |\phi|^4 +\cdots
\nonumber \\ & \quad %
- (\phi J^\dag+\phi^\dag J)~, 
\\
\frac{\Gamma}{2} &= \frac{1+3i}{5} \frac{\left(\frac{\rh}{L}\right)^{\zl}}{(eL)^2}~.
%
\end{align}
\end{subequations}

\end{itemize}

\end{document}